\documentclass[ onecolumn, prd, showpacs]{revtex4}

\usepackage{bm}
\usepackage{graphicx}
\usepackage{psfrag}

\newcommand{\dd}{\ensuremath{\textrm{d}}}
\newcommand{\Ee}{\ensuremath{\textbf{E}^3}}

\newcommand{\UD}[2]{\ensuremath{^{#1}_{\phantom{#1} #2}}}

\newcommand{\pard}[2]{\ensuremath{\frac{\partial #1}{\partial #2}}}

\newcommand{\tDd}{\ensuremath{ \widetilde{\cal D}}}
\newcommand{\Dd}{\ensuremath{ {\cal D} }}
\newcommand{\Nn}[1]{\ensuremath{ {\cal N}\left[ {#1} \right] }}
\newcommand{\Nnn}{\ensuremath{ {\cal N} }}
\newcommand{\Pp}{\ensuremath{ {\cal P} }}

\newcommand{\Ppm}[1]{\ensuremath{ {\cal P}^{-1}\left[ {#1} \right] }}
\newcommand{\Ppmm}{\ensuremath{ {\cal P}^{-1}}}
\newcommand{\NPm}[1]{\ensuremath{ {\cal NP}^{-1}\left[ {#1} \right] }}

\newcommand{\dott}[1]{\ensuremath{{#1}\dot{}}}

\newcommand{\av}[1]{\ensuremath{\langle{#1}\rangle}}

\newcommand{\leftsym}[1]{\ensuremath{\left({#1}\right.}}
\newcommand{\rightsym}[1]{\ensuremath{\left.{#1}\right)}}
\newcommand{\leftasym}[1]{\ensuremath{\left[ {#1} \right.}}
\newcommand{\rightasym}[1]{\ensuremath{\left. {#1} \right] }}

\newtheorem{theorem}{Theorem}[section]

\newenvironment{remark}[1][Remark.]{\begin{trivlist}
\item[\hskip \labelsep {\bfseries #1}]}{\end{trivlist}}

\newcommand{\qed}{\nobreak \ifvmode \relax \else
      \ifdim\lastskip<1.5em \hskip-\lastskip
      \hskip1.5em plus0em minus0.5em \fi \nobreak
      \vrule height0.75em width0.5em depth0.25em\fi}

\begin{document}
\title{Covariant coarse--graining of inhomogeneous dust flow in General Relativity}
\author{Miko\l{}aj Korzy\'nski}
\affiliation{Max--Planck--Institut f\"ur Gravitationsphysik \\(Albert--Einstein--Institut)\\
Am M\"uhlenberg 1, 14476 Potsdam/Golm, Germany}
\email{mikolaj.korzynski@aei.mpg.de}

\begin{abstract}
A new definition of coarse--grained quantities describing the dust flow in General Relativity is proposed. It assigns the coarse--grained expansion, shear and vorticity to
finite--size comoving domains of fluid in a covariant, coordinate--independent manner. The coarse--grained quantities are all quasi--local functionals, depending only on the geometry of the boundary of the considered domain. 
They can be thought of as relativistic generalizations of simple volume averages of local quantities in a flat space. The procedure is based on the isometric embedding theorem for $S^2$ surfaces and thus requires the boundary of the domain in question to have spherical topology and positive scalar curvature. We prove that in the limit of infinitesimally small volume the proposed quantities reproduce the local
expansion, shear and vorticity.
In case of irrotational flow we derive the time evolution equation for the coarse--grained quantities and show that its structure is very similar to the evolution equation for their local counterparts. Additional terms appearing in it may serve as a measure of
the backreacton of small--scale inhomogeneities of the flow on the large--scale motion of the fluid inside the domain
and therefore the result may be interesting in the context of the cosmological backreaction problem.  
We also consider the application of the proposed coarse--graining procedure to a number of known exact solutions of Einstein equations with dust and show that it yields reasonable results.

\end{abstract} 

\maketitle

\section{Introduction}

During the last few years there has been a sudden growth of interest in the problem of averaging of the Einstein equations of gravitation. The topic attracted the attention 
of the relativist community because the influence of small--scale inhomogeneities in the Universe upon its large--scale behavior, neglected by most cosmologists so far, may possibly offer a simple and attractive explanation of the accelerated expansion of the Universe (\cite{Wiltshire:2007jk, wiltshire-2008-17, Wiltshire:2007fg, rasanen-2006-15, rasanen-2004-0402}, though see also \cite{Ishibashi:2005sj} and \cite{Paranjape:2008jc} for a different view). The discussion seems inconclusive so far, but apart from the application to cosmology the problem of averaging the Einstein
equations is very interesting by itself, as it touches upon a fundamental question concerning the application of Einstein's equation to physical and astrophysical situations. Namely, when we apply Einstein's theory to objects like galaxies or galaxy clusters, often treating the stars, black holes and other bodies of which they are composed as dust or fluid,  we also effectively replace a complicated metric, describing the fine details of spacetime in the vicinity of those objects, by a smooth one. When one realizes that, the fundamental questions of validity of such an approximation and corrections one should possibly include arise immediately \cite{zalaletdinov-1, ellis-larger}. 

Up to now the study of cosmological backreaction has mostly proceeded along two main lines of research. The first is the study of several families
of exact, non--homogeneous solutions of Einstein's equations filled with dust of fluid (Lema\^itre--Tolman--Bondi \cite{Celerier:1999hp, Tomita:2000jj, Bolejko:2005fp, Celerier:2009sv}, Swiss cheese \cite{Marra:2007pm, Valkenburg:2009iw, Kozaki:2002ka}, Szekeres \cite{Bolejko:2005fp} etc.). The second one consists of attempts to derive a general expression for backreaction in a non--homogeneous metric,
using perturbation expansion  and/or a coarse--graining (``averaging'') scheme \cite{marochnik-1980, isaacson-1968, Clarkson:2009hr, Larena:2009md, Behrend:2007mf, Brown:2009tg}. In this paper we follow the second line and suggest a new approach to deriving an exact, non--perturbative formula for backreaction.
 
The averaging scheme for the metric and matter fields is a crucial element of any backreaction study. It allows for smoothing the fine structure
of a given spacetime and matter fields, pushing the entire configuration of fields  towards a more symmetric and homogeneous one. In statistical physics the procedure of forgetting the
fine, small--scale structure of a physical system is known as \emph{coarse--graining}. \footnote{The coarse--grained quantities are very often averages of the original ones, but this is by no means an absolute requirement; for that reason I prefer to use the term coarse--graining over more popular averaging.} 
 
Clearly there are infinitely many ways to perform coarse--graining and the approach one chooses must be adapted to the physical system under consideration. In the context of cosmology, or broadly speaking General Relativity,
several schemes have been proposed in the past.
The most popular one is the Buchert's scheme \cite{Buchert:1999mc}, in which one assigns scalar cosmological parameters -- the volume expansion, the average scalar curvature and matter density -- to a comoving domain in 3+1 splitting, using its total volume and volume averaging. The resulting formalism is quite simple, but
it suffers from a fundamental drawback -- it only allows for coarse--graining of the scalar part of the evolution equations, 
completely ignoring the tensorial part. Zalaletdinov \cite{zalaletdinov-1, zalaletdinov-2}, proposed a framework based on a globally defined field of ``bivectors'', connecting
tangent spaces of distant points in spacetime. It is more general than Buchert's and allows for coarse--graining of the full Einstein equations, but it requires an
earlier choice of the bivectors, which introduces certain unwanted arbitrariness into the results. It was later developed and applied to the cosmological context by Coley, Pelavas and Zalaletdinov \cite{coley-2005-95, coley-2007-75}, as well as Singh and Paranjape \cite{Paranjape:2008mx}. Other approaches have been proposed by Anastopoulos \cite{anastopoulos-2009-79}, Carfora and Marzuoli \cite{carfora-1984}, based on the Ricci flow deformation, later developed in \cite{carfora-1995, buchert-2002, Buchert:2002ht}, and Reiris \cite{Reiris:2007gb}. 

In this paper we will propose an alternative to Buchert's and other coarse--graining schemes. We will present
a procedure which assigns the coarse--grained expansion, shear and vorticity to a finite--sized comoving domain of
the fluid in a coordinate--independent manner. The proposed quantities are quasi--local, \emph{i.e.} they are functionals
of the geometry of the \emph{boundary} of the coarse--graining domain. While quasi--local quantities are
abundant in mathematical relativity \cite{lrr-2009-4}, they haven't been used too much in this particular context (though see \cite{Sussman:2008vs}, \cite{Afshar:2009pi}). 
The shear and  expansion are functionals of the induced metric on the boundary and its time derivative, while the definition
of vorticity involves the shift vector field at the boundary. For computational simplicity we develop our formalism  with pressureless dust as the only matter field present, but the construction carries over easily to more complicated fluids.

The definitions proposed here are generalizations of the volume averages of shear, expansion and vorticity in non--relativistic, Newtonian theory of gravitating fluids. The basic idea is the following:
we embed the boundary of a comoving domain isometrically into the three--dimensional Euclidean space and construct a 
fictitious, three--dimensional fluid velocity field on the image of the embedding. The velocity field is chosen in such a way that its `fictitious' flow induces the same infinitesimal metric deformation on the embedded surface as the `true' dust flow does on the domain boundary in the spacetime. This fictitous velocity is then used to assign the coarse--grained expansion
and shear via the standard, ``Newtonian'' volume average of the velocity gradient, transformed into a boundary integral. The vorticity on the other hand cannot be 
read out from the boundary metric only, so in order to define it we use a different approach, involving the 
pushforward of the ADM shift vector to the Euclidean space. In both cases the construction and the uniqueness of results rely on the existence and rigidity of the isometric embedding, guaranteed by the well--known theorem conjectured by Weyl and established by Lewy, Nirenberg, Heinz, Pogorelov, Alexandrov 
 and Cohn--Vossen (see \cite{spivak-chapters, han-hong} and references therein). The embedding theorem imposes two
 relatively 
mild restrictions on the coarse--graining domain, namely its boundary must be homeomorphic to a two--sphere and must have a positive curvature. The first one is quite harmless, as most interesting domains are ball--shaped anyway and thus have a topologically spherical boundary. The second should be satisfied if the domain is ``round enough'', \emph{i.e.} its shape is close to a geodesic ball or ellipsoid. 

The strength of the approach presented in this paper lies mainly in the fact that it is manifestly covariant, as each step can be defined in a geometric, coordinate--independent manner. Apart from  the domain itself and the 3+1 splitting of the spacetime, one does not have to make any additional, artificial choices which might influence the final result, unlike the approaches of Anastopoulos or Zalaletdinov. 
A clever use of the embedding theorem allows one to bypass the problems of covariant averaging of tensor fields and consequently the resulting decomposition of the flow into the large--scale, coarse--grained part and local inhomogeneities is covariant. It is also worth mentioning that the expression for backreaction one finally obtains has the form of a surface integral divided by volume. This makes it relatively easy to estimate its magnitude given estimates on the magnitude of local inhomogeneities.

The main purpose of coarse--graining over a domain is to find a best fitting homogeneous cosmological solution. In Buchert's original scheme the fitting is performed to a FLRW metric via three scalar parameters, one of them being the suitably defined expansion. The advantage of including the tensorial shear and vorticity is that it expands the class of homogeneous solutions one might want to fit to non--isotropic ones. 
The presented procedure is more mathematically involved than Buchert's, but still it seems to be a good compromise between mathematical elegance and computational 
simplicity, as it only requires solving two systems of well--known PDE's and evaluating a surface integral.

The paper is organized as follows. In the next section we will briefly discuss the theory of Newtonian gravitating dust
(``Newtonian cosmology''), the coarse--graining of its evolution equations and discuss the backreaction. The section is intended to be a short review and summary of the most important results, for an exhaustive treatment of the subject see \cite{buchert-1997-320, ehlers-1997-29} and \cite{Zalaletdinov:2002sv, Zalaletdinov:2002sw, Zalaletdinov:2002sx}. It is included to provide the motivation for the relativistic coarse--graining procedure.

In the third section we will spell out the definitions of coarse--grained expansion, shear and vorticity for any spacetime filled with dust as well as the coarse--grained counterpart of the Fermi derivative operator for irrotational dust. In the fourth one we will prove that the coarse--grained quantities coincide with the local ones if the coarse--graining domain shrinks to 
a point, thus showing that the definitions are reasonable. In the fifth section we will derive the time evolution equation for the coarse--grained quantities for irrotational dust and show that it has a similar structure to the local one. We will derive thereby an exact, non--perturbative expression for backreaction in a spacetime filled by non--rotating dust. The rotational case will be discussed in the next paper.

Finally we will discuss briefly the application of the presented formalism to several known exact
cosmological solution of Einstein's equations.

\section{Newtonian cosmology}

Consider a cloud of dust in ordinary Euclidean space $\Ee$, interacting via Newtonian gravity. In Cartesian coordinates $(x^a)$ it is described by the local density function $\rho(x^a,t)$,
velocity field $v^a(x^b,t)$ and the Newtonian potential $\phi(x^a,t)$. They satisfy a well--known system of PDE's
\begin{eqnarray}
 \pard{v^a}{t} + v^b\,\pard{v^a}{x^b} &=& -\phi^{,a} \label{eqvelocity}\\
 \pard{\rho}{t} + v^a\,\pard{\rho}{x^a} &=& -\rho\,\pard{v^a}{x^a} \label{eqrho}\\
 \Delta \phi = 4\pi G\,\rho \label{eqpoisson}.
\end{eqnarray}
The first one is the equation of motion of the dust particles in the gravitational field, the second is the continuity equation and the third is the Poisson equation for the Newtonian potential.
The left hand side of equations (\ref{eqvelocity}) and (\ref{eqrho}) is usually called the
convective derivative and denoted by $\frac{D v^a}{\partial t}$ and $\frac{D\rho}{\partial t}$ respectively.

Consider the solutions to (\ref{eqvelocity})--(\ref{eqpoisson}) of the form
\begin{eqnarray*}
 v^a(x^b,t) &=& Q\UD{a}{b}(t)\,x^b \\
 \rho(x^a,t) &=& \rho (t) \\
 \phi(x^a,t) &=& \frac{1}{2}\,\Phi_{ab}(t)\,x^a\,x^b.
\end{eqnarray*}
Solutions of this type are homogeneous in the sense that the matter density is constant in space and the matter
undergoes a homogeneous expansion, stretching and rotation. 
The ansatz plugged into the equations of motion it yields a system of non--linear ODE's
\begin{eqnarray}
 \dot Q\UD{a}{b} &=& -Q\UD{a}{c}\,Q\UD{c}{b} - \Phi\UD{a}{b} \label{eqvelocity2}\\
 \dot \rho &=& -\rho \,Q\UD{a}{a} \label{eqrho2} \\
 \Phi\UD{a}{a} &=& 4\pi G\,\rho \label{eqpoisson2}.
\end{eqnarray}
for one general matrix $Q\UD{a}{b}$, a symmetric matrix $\Phi_{ab}$ and a scalar function $\rho$. 
$Q\UD{a}{b}$ can be decomposed into the trace, called expansion, the symmetric tracesless part (shear) and the antisymmetric part (vorticity)
\begin{eqnarray*}
 Q_{ab} = \frac{1}{3}\,\theta\,\delta_{ab} + \sigma_{(ab)} + \omega_{[ab]}.
\end{eqnarray*}
Similar procedure applied to $\Phi_{ab}$ yields the scalar part $\Phi$, directly related to $\rho$ via (\ref{eqpoisson2}), and the traceless part 
\begin{eqnarray*} 
\Phi_{ab} = \frac{1}{3}\,\Phi\,\delta_{ab} + \Sigma_{(ab)}.
\end{eqnarray*}
$\Sigma_{ab}(t)$ can be interpreted as the large--scale tidal field. It doesnt't have any evolution equations of its own, which makes the system (\ref{eqvelocity2})--(\ref{eqpoisson2}) underdetermined. We can close it up if we assume the tidal field to vanish.

The homogeneous solutions discussed here contain as a special case the Newtonian counterparts of FLRW solutions. It suffices to assume that
the flow is rotation--free and homogeneous ($Q_{ab} = H\,\delta_{ab}$) and equations (\ref{eqvelocity2})--({\ref{eqpoisson2}) yield the standard Friedmann equations with dust matter.

\subsection{Coarse--graining of the Newtonian cosmology}

In order to calculate backreaction  we need to split all relevant quantities into the large--scale, coarse--grained part and small--scale inhomogeneities.
Fix a compact domain  $G_t \in \Ee$, dragged by the dust particles. We assign coarse--grained quantities to $G_t$ by simple volume averaging
\begin{eqnarray}
 \av{Q\UD{a}{b}} &=& \frac{1}{V}\,\int_{G_t} v\UD{a}{,b}\,\dd^3 x \label{eqdefQnewton}\\
 \av{\phi_{,ab}} &=& \frac{1}{V}\,\int_{G_t} \phi_{,ab}\,\dd^3 x \label{eqdefpot}\\
 \av{\rho} &=& \frac{1}{V}\,\int_{G_t} \rho\,\dd^3 x \label{eqdefrho} ,
\end{eqnarray}
where $V$ denotes the time--dependent volume of $G_t$.

Note that by the virtue of the divergence theorem the first two averages are effectively \emph{surface integrals}:
\begin{eqnarray}
\av{Q\UD{a}{b}} &=& \frac{1}{V}\,\int_{\partial G_t} v^a\,n_b\,\dd\sigma \label{eqQbdr}\\
\av{\phi_{,ab}} &=& \frac{1}{V}\,\int_{\partial G_t} \phi_{,a}\,n_b\,\dd\sigma \nonumber.
\end{eqnarray}
They are therefore only sensitive to the values of $v^a$ and $\phi_{,b}$ at the boundary of the coarse--graining domain. We can now split the velocity, density and the Newtonian potential into the large--scale part and the small--scale inhomogeneities:
\begin{eqnarray*}
 v^a &=& \av{Q\UD{a}{b}}\,x^b + \delta v^a \\
 \phi &=& \frac{1}{2}\,\av{\phi_{,ab}}\,x^a\,x^b + \delta \phi \\
 \rho &=& \av{\rho} + \delta \rho.
\end{eqnarray*}
Now   the problem of backreaction in Newtonian cosmology can be formulated as a precise mathematical question: how do the evolution equations for the coarse--grained quantities differ from those obtained with the assumption of complete homogeneity, \emph{i.e.} (\ref{eqvelocity2})--(\ref{eqpoisson2})?

It turns out that despite the nonlinearity of the system (\ref{eqvelocity})--(\ref{eqpoisson}) the former  have the same structure as the latter ones
\begin{eqnarray}
  \dott{\av{Q\UD{a}{b}}} &=& -\av{Q\UD{a}{c}}\,\av{Q\UD{c}{b}} - \av{\phi\UD{,a}{,b}} + B\UD{a}{b} \label{eqvelocity3}\\
  \dott{\av{\rho}} &=& -\av{\rho}\,\av{Q\UD{a}{a}} \label{eqrho3}\\
  \av{\phi\UD{,c}{,c}} &=& 4\pi G\,\av{\rho} \label{eqpoisson3}
\end{eqnarray}
apart from just one additional term in the first equation:
\begin{eqnarray}
B\UD{a}{b} = V^{-1}\,\int_{\partial G_t} (\delta v^c\,\delta v\UD{a}{,b}\,n_c - \delta v^c\,\delta v\UD{a}{,c} \,n_b)\,\dd\sigma. \label{eqnewtonbackreaction}
\end{eqnarray}
Clearly this is a backreaction term, as it describes the influence of the inhomogeneities upon the dynamics of the large--scale quantities. Equation (\ref{eqnewtonbackreaction}) was first derived in the special case of $Q\UD{a}{b} = H\,\delta\UD{a}{b}$ by Ehlers and Buchert \cite{buchert-1997-320}.

Expression (\ref{eqnewtonbackreaction}) for backreaction has a number of interesting features. First, note that it depends on velocity inhomogeneities, but not on the density inhomogeneities. Consequently the backreaction is exactly the same for a smooth matter distribution and a collection of point particles. 
The physical reason behind this surprising fact is the equivalence principle: in absence of pressure the motion of a particle is only affected by gravitational forces, which in turn does not depend on its mass. Therefore $\rho$ drops out of equation (\ref{eqvelocity}) and $\delta \rho$ does not appear in the expression for backreaction. 

Secondly, $B_{ab}$ depends only on the values of $\delta v^a$ at $\partial G_t$. Although (\ref{eqnewtonbackreaction}) appears to involve its derivatives
in all directions,  the normal derivative drops out because of antisymmetrization in $b$ and $c$.

Finally equation (\ref{eqnewtonbackreaction}) has the form of a surface integral divided by the enclosed volume. This is of course a pleasant consequence 
of the fact that the expressions for $\av{Q\UD{a}{b}}$ and $\av{\phi_{,ab}}$ are effectively surface integrals too. Thus $\dott{\av{Q\UD{a}{b}}}$ 
depends only on the dynamics of particles at the boundary of the domain and any backreaction terms arising in that equation must be surface expressions too.

This form makes equation (\ref{eqnewtonbackreaction})  a very convenient tool to provide bounds on backreaction on very large scales. Consider a ball--shaped
domain of radius $R$. Integrating over the surface yields the factor of $R^2$, while by dividing by volume amounts to multiplying by $R^{-3}$. If we can give absolute bounds on $\delta v^a$ and their derivatives on the spacetime, we can obtain even stronger bounds on the backreaction of the form of
\begin{eqnarray*}
 |B\UD{a}{b}| < \frac{C}{R}
\end{eqnarray*}
for a positive constant $C$.

Let us stress that the coarse--grained equations (\ref{eqvelocity3})--(\ref{eqpoisson3}) are \emph{exact}, they do not involve any kind approximation or perturbation expansion. In fact, (\ref{eqvelocity3})--(\ref{eqpoisson3}) are just the rewriting of (\ref{eqvelocity})--(\ref{eqpoisson}) in terms of non--local variables (\ref{eqdefQnewton})--(\ref{eqdefrho}) and they hold universally for any solution. In particular, expression (\ref{eqnewtonbackreaction}) yields the
correct value of backreaction no matter how much the solution considered differs from the corresponding homogeneous one.

\section{Coarse--graining in General Relativity}

The result for Newtonian backreaction, obtained in the previous section, suggests a new, promising way to approach the backreaction problem in the relativistic setting: one could derive the relativistic counterpart of equation (\ref{eqvelocity3}) for the time derivative of a non--local, coarse--grained velocity gradient, involving additional terms coming from the metric and flow inhomogeneities.

Consider a curved spacetime filled with gravitating dust, described by energy density $\rho$ and four--velocity field $u^\mu$.
Define the gradient of $u^\mu$ 
\begin{eqnarray}
 Z_{\mu\nu} = \nabla_\nu u_\mu \label{eqZDU}, 
\end{eqnarray}
which is completely orthogonal to $u^\mu$
\begin{eqnarray*}
 Z_{\mu\nu} \,u^\mu = Z_{\nu\mu} u^\mu = 0.
\end{eqnarray*}
 This makes $Z_{\mu\nu}$ effectively a three--dimensional tensor whose role is analogous to  $v\UD{a}{,b}$ from the previous section. It is conveniently decomposed into the \emph{local} expansion, shear and vorticity
\begin{eqnarray*}
 Z_{\mu\nu} &=& \frac{1}{3}\theta\,h_{\mu\nu} + \sigma_{(\mu\nu)} + \omega_{[\mu\nu]} \\
h_{\mu\nu} &=& g_{\mu\nu} + u_\mu\,u_\nu
\end{eqnarray*}
which describe how an \emph{infinitesimal} fluid or dust element is deformed and how it rotates during its motion.
$Z_{\mu\nu}$ satisfies an evolution equation along the worldlines of dust particles
\begin{eqnarray}
 \nabla_u Z\UD{\mu}{\nu} = -Z\UD{\mu}{\rho}\,Z\UD{\rho}{\nu} - R\UD{\mu}{\rho\nu\sigma}\,u^\rho\,u^\sigma.\label{eqduZ}
\end{eqnarray}
This equation is clearly analogous to (\ref{eqvelocity2}), since the contraction of the Riemann tensor on the right hand side can be decomposed into the trace, related to the Einstein tensor and thus to the local matter content, and the traceless Weyl tensor describing the tidal forces, just like $\Phi\UD{a}{b}$ in (\ref{eqvelocity2}).
The trace of equation (\ref{eqduZ}) is  the well--known Raychaudhuri equation.

Let us try to push the analogy with Newtonian cosmology further and consider a \emph{finite} fluid element traveling through spacetime. All particles enclosed within it make up a four--dimensional cylinder $C$ in spacetime and particles contained in its boundary generate a three--dimensional tube $\partial C$ (see fig. \ref{figtube}). The tube can be then foliated by previously chosen constant time slices. One could then propose a way of assigning the coarse--grained $Z_{\mu\nu}$ (and thus the expansion, shear and vorticity) to the constant time slices of $C$. The  interior of these slices $C_t$ would play the role of the comoving coarse--graining domain. We could then derive the time evolution equation for the coarse--grained $\av{Z_{\mu\nu}}$ and,  if the proposed definition is appropriate, the evolution equation should have a similar form to (\ref{eqduZ}) with a bunch of additional terms.  These new terms should arise from the inhomogeneities of the metric and the dust flow  inside and on the boundary of  the domain. They would describe how these small--scale inhomogeneities influence the large--scale motion of the coarse--graining domain as a whole. Like $B_{ab}$ in (\ref{eqvelocity3}), they should  vanish on a homogeneous FLRW solution but not in a general
situation, which would justify referring to them as the backreaction terms.

\begin{figure}
 \includegraphics[scale=0.7]{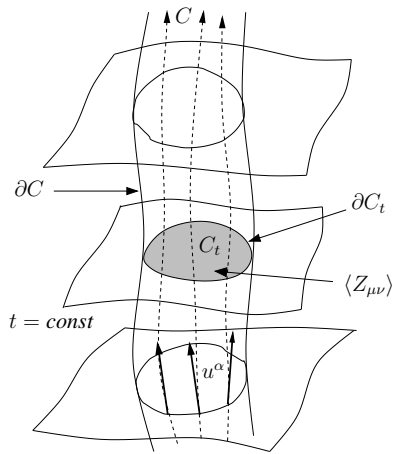}
 \caption{Four--dimensional cylinder $C$, generated by particles within a finite volume of the fluid, and its boundary $\partial C$, a three--dimensional tube of worldlines, both foliated by constant time slices.}
\label{figtube}
\end{figure}

Obviously there are infinitely many possible ways to assign $\av{Z_{\mu\nu}}$  to a finite domain. We should however
make sure that the prescription satisfies a couple of reasonability conditions.
First, we the new definition should be consistent with (\ref{eqZDU}), \emph{i.e.} we should recover the local velocity gradient $\nabla_\nu u_\mu$ if we shrink the tube under consideration to an infinitesimally small one. 
Second, a good coarse--graining prescription should be covariant in the following sense:  apart from the choice of the tube itself and the 3+1 splitting of the spacetime, the result \emph{should not depend upon any externally introduced structure, including the coordinate system}. This requirement is not just 
a question of aesthetics or mathematical pedantry, as any dependence of this kind introduces spurious degrees of freedom which must be separated out carefully to obtain 
meaningful, physical results.

While scalar quantities like expansion are easy to coarse--grain  in a covariant way by taking the volume average, no simple way to coarse--grain tensors exists in a general, curved spacetime.  In fact, the covariance
problem appears here at an even earlier stage: note that the coarse--grained vectors and tensors cannot live on their own, they need a vector space to be defined at all. A sensible
prescription must begin by assigning a vector space to each coarse--graining domain, once again in a coordinate--independent way.

We will now propose a relatively simple procedure of assigning  $\av{Z_{\mu\nu}}$ to finite elements of the fluid, which can be proven to work under relatively mild conditions. We will prove that it satisfies the requirements  stated above and derive the evolution equation for  $\av{Z_{\mu\nu}}$ in the special
case of irrotational dust flow. More general case will be treated in the next papers.

The main tool we are going to use is the isometric embedding theorem for $S^2$ surfaces: 
\begin{theorem}[Isometric embedding theorem for $S^2$]
 Given a compact, orientable surface $S$ homeomorphic to $S^2$, with positive metric $q$ whose scalar
curvature $R > 0$. Then 
\begin{itemize}
 \item there exists an isometric embedding
\begin{eqnarray*}
 f: S \mapsto \Ee 
\end{eqnarray*}
into the 3--dimensional Euclidean space;
\item the embedding is unique up to rigid rotations, translations and reflexions.
\end{itemize}\label{thiso}
\end{theorem}
(see \cite{spivak-chapters} for an exhaustive treatment of the topic and the history of developments in this area; in particular the existence part is discussed in  \cite{spivak-embedding-existence}, and the uniqueness in \cite{spivak-cohn-vossen}).
If we fix the orientation, the theorem states simply that a compact, two--dimensional surface of $S^2$ topology, satisfying the positive curvature condition, can be recognized as a surface in $\Ee$ and moreover
this can be done in only one way, up to moving the surface around or rotating it as a whole.
Let us stress that the hypotheses of the isometric embedding theorem cannot be relaxed, the theorem is in general untrue if they are violated. Compact, orientable surfaces which satisfy them will be called \emph{admissible}.

The isometric embedding theorem allows one to recognize the boundary of the coarse--graining region $\partial C_t$ as a submanifold in $\Ee$ as long as it remains admissible. $\partial C_t$ is at the same time the constant time section of the tube $\partial C$. 
We will show that one can read out $\av{Z_{\mu\nu}}$ \emph{just from the geometric data on $\partial C_t$, completely ignoring
the geometry inside the coarse--graining domain}. While the idea of assigning a coarse--grained quantity to a finite volume in a quasi--local manner, \emph{i.e.} using just the geometry of its two--dimensional boundary, may seem strange at first, note that this strategy worked remarkably well in the Newtonian case: (\ref{eqdefQnewton}) depends effectively only on the \emph{surface} values of $v^a$, the behavior of the velocity field inside the domain simply does not influence the its value. 

\subsection{Notation, 3+1 splitting, conventions regarding the indices}

In this paper we will mostly work in a comoving, though not necessary orthogonal coordinate system $(t,y^i)$, \emph{i.e.} in coordinates in which the following conditions
\begin{eqnarray*}
 u^\mu\,y\UD{i}{,\mu} &=& 0 \\
 u^\mu\,t_{,\mu} &=& 1
\end{eqnarray*}
hold. A coordinate system of this kind should exist at least locally. Its constant time surfaces are spacelike and will be denoted by $\Sigma_t$, while their orthonormal, future--pointing vector will be referred to as $\nu^\alpha$.
We can perform the standard ADM decomposition of the metric and curvature in this coordinate system, obtaining
a positive definite 3--metric $h_{ij}$, laps function $N$ and shift vector $N^i$, orthogonal projection 
operator $P\UD{\alpha}{\beta} = \delta\UD{\alpha}{\beta} + \nu^\alpha\,\nu_{\beta}$ and the extrinsic curvature
$K\UD{i}{j} = -P\UD{i}{\alpha}\,P\UD{\beta}{j}\,\nabla_{\beta}\,\nu^\alpha$. The covariant derivative operator on $\Sigma_t$, adapted to $h_{ij}$, will be denoted by $D$.
  
In comoving coordinates the worldlines of all dust particles are the curves of constant spatial coordinates $y^i$. We can therefore describe 
any tube $\partial C$ of particle worldlines by equation of the type $y^i = \xi^i(\theta^A)$, where $\theta^A$ are two coordinates on the tube sections and $\xi^i(\theta^A)$ is a set of three functions.
The tube itself is now parameterized by the coordinate time $t$ and $\theta^A$.

The constant time sections of the tube will be denoted by $\partial C_t$ and the domain of $\Sigma_t$ enclosed by them by $C_t$. 
The metric induced on $\partial C_t$, expressed in $(\theta^A)$, is given by
\begin{eqnarray}
 q_{AB}(t,\theta^B) = \xi\UD{i}{,A}\,\xi\UD{j}{,B}\,h_{ij}. \label{eqqAB}
\end{eqnarray}

As usual, the Greek indices ($\mu$, $\nu$, ...) run from 0 to 3 and will be used with geometric objects defined on the tangent space to the spacetime. The lower case Latin letters $i,j,...$ are assumed to run from 1 to 3 and will denote
objects on the space tangent to constant time hypersurfaces $\Sigma_t$. The lower case letters from the beginning of the alphabet ($a,b,c,...$) also run from 1 to 3, but will be used with objects defined on the Euclidian space $\Ee$.
Finally the upper case Latin indices $A, B, ...$ run form 1 to 2 and will be used with objects on two--dimensional boundaries of coarse--graining domains (constant time sections of the worldline tubes).
  
\subsection{The coarse--graining procedure}

Let $\partial C$ denote a worldline tube whose constant time sections $\partial C_t$ are admissible in the sense of theorem \ref{thiso}.
Consider a family of time--dependent isometric embeddings $$f_t: \partial C_t \mapsto \partial D_t \subset \Ee,$$
where $\partial D_t$ is the image of $f_t$.
Let $x^a$, $a=1...3$ be the  orthogonal, Cartesian coordinates in $\Ee$. The embeddings can be described
by equations
\begin{eqnarray}
 x^a = \chi^a(t,\theta^A) \label{eqchi}
\end{eqnarray}
with three functions $\chi^a(t,\theta^A)$ satisfying the isometry condition
\begin{eqnarray}
 q_{AB}(t,\theta^A) = \delta_{ab}\,\chi\UD{a}{,A}\,\chi\UD{b}{,B} \label{eqisometry}.
\end{eqnarray}
(\ref{eqisometry}) is a non--linear partial differential equations system and while theorem \ref{thiso} guarantees the existence of solutions, they are in general impossible to find analytically. However numerical schemes for solving it have been developed in the context of black hole theory \cite{bondarescu-2002-19, nollert}. 
 
The isometric embeddings can be freely rotated and moved around in $\Ee$ without violating (\ref{eqisometry})
\begin{eqnarray}
 \chi^a(t,\theta^A) \to R\UD{a}{b}(t)\,\chi^b(t,\theta^A) + W^{a}(t).
\end{eqnarray}
For any choice of the isometric embeddings we can follow the image of each individual particle of $\partial C$, labeled by coordinates $\theta^A$,
in the Euclidean space, obtaining thereby its fictitious trajectory in $\Ee$. If the trajectory is regular enough, we can also assign to it a fictitious instantaneous velocity $v^a$ at each time (fig. \ref{figembs}). 
\begin{figure}
 \includegraphics[width=8cm]{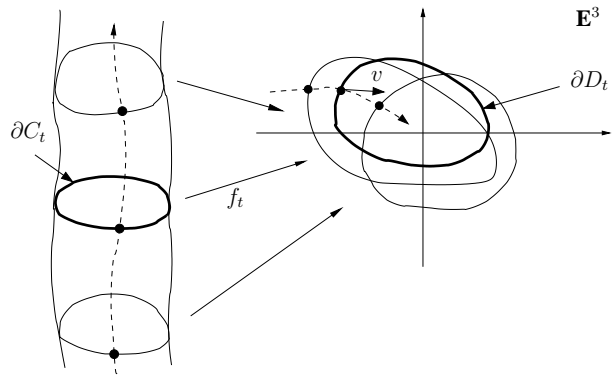}
 \caption{One--parameter sequence of embeddings and the fictitious Euclidean trajectory of a single particle}
 \label{figembs}
\end{figure}

In fact, we do not need to create a whole family of embeddings to recover $v^a$ of each particle. For a given instance of time it is enough to have one embedding and the time derivative
of the induced metric in comoving coordinates. This is guaranteed by the linearized version of the embedding theorem:
 \begin{theorem}
  Given a compact, orientable surface $S$ homeomorphic to $S^2$, embedded isometrically into $\Ee$, whose scalar curvature $R>0$,
  and a symmetric tensor field $r_{AB}$ on $S$. Then
 \begin{itemize}
  \item there exists a vector field $v^a$, $v^a(x^b) \in \Ee$, defined on $S \subset \Ee$, such that
       \begin{eqnarray*}
        \dot q_{AB} = r_{AB}
       \end{eqnarray*}
	when $S$ dragged along $v^a$,
  \item $v^a$ is unique up to adding a vector field $Y^a$ of the form
    \begin{eqnarray}
     Y^a &=& \Omega^a_{\phantom{a}b}\,x^b + W^a \label{eqambi}\\
     \Omega_{ab} &=& -\Omega_{ba} = \textrm{const}, \quad W^a = \textrm{const}.\nonumber
    \end{eqnarray}
 \end{itemize}\label{thiso2}
 \end{theorem}
In short, the theorem states that if we are given a single embedding of a surface and a time derivative of the metric in form of a symmetric tensor, we can always find the three--velocity field $v^a$, defined on the image of the surface, which induces the right metric deformation at the linear order when the surface is dragged along it. 
The vector field is unique up to a rotational and translational part, which constitute a six--dimensional vector space. Let us stress that $v^a$ is defined \emph{only at $\partial D_t$}, our procedure \emph{does not} assign velocity vectors to points inside or outside the image of the embedding (fig. \ref{figv}).

\begin{figure}
 \includegraphics[width=9cm]{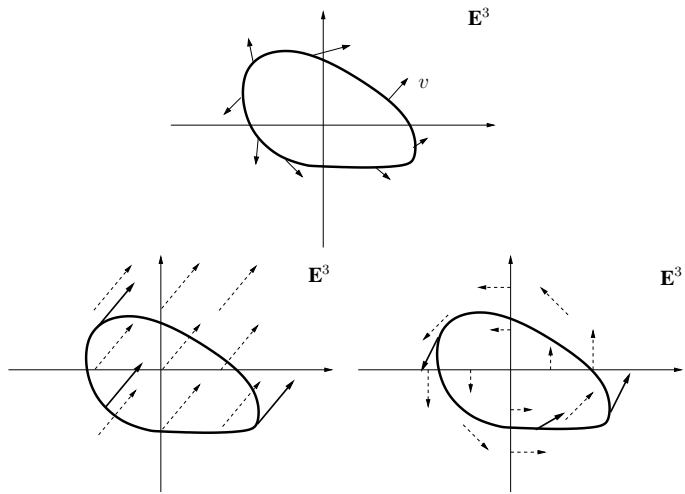}
 \caption{Vector field, defined on a convex surface, inducing given metric change, is unique up to the restrictions
of rotational and translational vector fields to that surface}
 \label{figv}
\end{figure}

Theorem \ref{thiso2} follows easily from \ref{thiso} if we consider a regular, one--parameter family of surfaces.
The existence and the regularity of the family of embeddings imply the existence of the time derivative of the embeddings with respect to the time parameter. This derivative yields the vector in question. The uniqueness of the
solution is guaranteed
by the second part of theorem \ref{thiso} (see also the ``infinitesimal rigidity'' theorem from \cite{spivak-inf-rigidity}).

 The vector field $v^a$ is related to $\dot q_{AB}$ via the action of a differential operator ${\cal P}$ given by
\begin{eqnarray}
 {\cal P}\left[v^a\right]_{AB} = 2v\UD{a}{\leftsym{,A}}\,\chi\UD{b}{\rightsym{,B}}\,\delta_{ab} \label{eqP1}
\end{eqnarray}
or equivalently, in a more geometric notation,
\begin{eqnarray}
  {\cal P}[v^a]_{AB} = 2v_{(A|B)} + 2v_n\,H_{AB} = \dot q_{AB} \label{eqP2}
\end{eqnarray}
where $v_A$ denotes the lower--index projection of $v$ to $\partial D_t$, $v_n$ is the outward normal part, the vertical line is the covariant derivative on $\partial D_t$ and $H_{AB}$ denotes its outward extrinsic curvature in $\Ee$. Theorem \ref{thiso2} can in fact be understood as a statement about the operator ${\cal P}$: it has a known, six--dimensional kernel of the form of (\ref{eqambi}) and a non--unique inverse ${\cal P}^{-1}$. If we constraint the domain of $\cal P$ by six independent, linear conditions on $v^a$, the restriction of ${\cal P}$ will be one--to--one and invertible.

We now apply the introduced machinery to our problem. For a given $t$ we obtain an embedding $f_t: \partial C_t \mapsto \partial D_t \subset \Ee$ and a vector field $v^a$ 
defined on $\partial D_t$. We can take the boundary expression (\ref{eqQbdr}) to define the symmetric part of $\av{Z_{ab}}$
\begin{eqnarray}
  \av{Z_{(ab)}} = \frac{1}{V_0}\,\int_{\partial D_t} v_{\leftsym{a}}\,n_{\rightsym{b}}\,\dd\sigma, \label{eqQsym}
\end{eqnarray}
where $V_0$ is the Euclidean volume of the domain $D_t$ enclosed by $\partial D_t$.
The key observation is that this symmetrized integral is insensitive to adding vector fields of type (\ref{eqambi}). Thus, irrespective of remaining ambiguity of  $v^a$,
(\ref{eqQsym}) defines unambiguously the expansion and shear of the flow, coarse--grained over the domain $C_t$, as tensors acting on the Euclidean space $\Ee$ \footnote{Alternatively one can think of them as of constant tensor fields on $\Ee$.}.

In order to coarse--grain the vorticity we follow a different strategy. First, note that in the comoving ADM coordinates the velocity gradient takes the form of
\begin{eqnarray*}
Z_{ij} = -N\,K_{(ij)} + D_{\leftsym{i}}N_{\rightsym{j}} - D_{\leftasym{i}}N_{\rightasym{j}}.
\end{eqnarray*}
(the minus sign in the last term is due to our convention concerning the order of indices in $Z_{ij}$).
We have shown how to extract the coarse--grained version of the first two terms from the induced metric $q_{AB}$. The antisymmetric part,
\emph{i.e.} the vorticity, does not influence $q_{AB}$. Note however that it depends only on the shift vector $N^i$. Therefore in order to define $\av{Z_{[ab]}}$, we will push $N^i$ directly
from the constant time slice $\Sigma_t$ to $\Ee$.
\begin{figure}
 \includegraphics[scale=0.8]{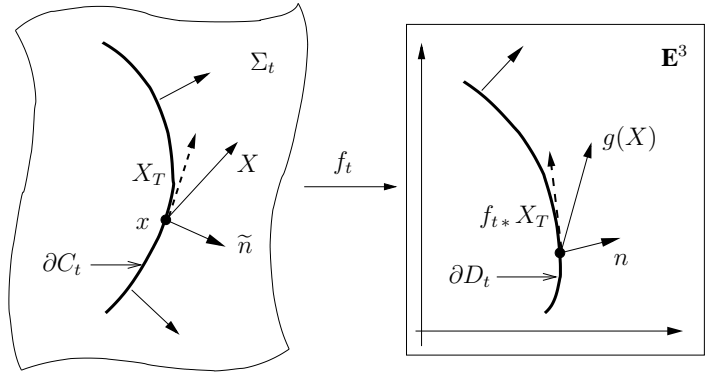} 
 \caption{Canonical isometry $g$ between the tangent space $T_x \Sigma_t$ and $\Ee$, induced by the isometric embedding}
 \label{figg}
\end{figure}

The pushforward is done via a canonical isometry $g$ between the tangent spaces $T_x \Sigma_t$, $x \in \partial C_t$, and $\Ee$, induced by the embedding $f_t$. It is defined as follows: 
we decompose any vector $X \in T_x \Sigma_t$ into the normal and tangent part with respect to $\partial C_t$. We then push the tangent part $X_T \in T_x \partial C_t$ by $f_{t\,*}$, while for the normal part we demand that
the product with the outward--pointing normal is the same in both spaces 
\begin{eqnarray*}
 g(X_T) &=& f_{t\,*}\,X_T \\
 g(X)^a\,n_a &=& X^i\,\tilde n_i.
\end{eqnarray*}
(see fig. \ref{figg}).
Note that $g$ can be used to push any geometric object, tensor or vector, from $T_x\Sigma_t$ to $\Ee$ and the other way round.
It is straightforward to check that $g$ is at each point $x\in \partial C_t$ an isometry between $T_x \Sigma_t$
and $\Ee$. We can now define vorticity  by
\begin{eqnarray}
 \av{Z_{[ab]}} = \frac{1}{V_0}\,\int_{\partial D_t} g(N)_{\leftasym{a}}\,n_{\rightasym{b}}\,\dd\sigma \label{eqQasym}.
\end{eqnarray}
It is possible and in fact convenient to use (\ref{eqQasym}) to fix the rotational part of $v^a$ by the condition
\begin{eqnarray}
 \int_{\partial D_t} v_{\leftasym{a}}\,n_{\rightasym{b}}\,\dd\sigma = \int_{\partial D_t} g(N)_{\leftasym{a}}\,n_{\rightasym{b}}\,\dd\sigma. \label{eqgauge}
\end{eqnarray}
 With this choice the whole $\av{Z_{ab}}$ is related to $v^a$ by equation
\begin{eqnarray}
\av{Z_{ab}} = \frac{1}{V_0}\int_{\partial D_t} v_a\,n_b\,\dd\sigma \label{eqZtot}
 \end{eqnarray}
and the analogy with (\ref{eqdefQnewton}) is complete. Moreover, if we choose the embedding $f_t$ at one time $t_0$, (\ref{eqgauge}) fixes them at all times up to irrelevant translations. These can be fixed as well, for example 
by demanding that the center of mass of $\partial D_t$  remains at the origin, but nothing in the presented the formalism depends on the postion of the surface within $\Ee$ or on the constant part of  $v^a$. Therefore the translational ambiguity of isometric embeddings is absolutely harmless.

For the sake of brevity we introduce the following notation for surface integrals of type (\ref{eqZtot}):
\begin{eqnarray*}
 \Nn{X_a}_b = \frac{1}{V_0}\,\int_{\partial D_t} X_a\,n_b\,\dd\sigma.
\end{eqnarray*}
We will also denote by ${\cal P}^{-1}\left[\,\cdot\,\right]$  the unique inverse of ${\cal P}$ satisfying
\begin{eqnarray*}
 \NPm{r_{AB}}_{[cd]} = 0
\end{eqnarray*}
plus an irrelevant condition fixing the constant part $W^a$.
With this notation the definition of the coarse--grained $Z_{ij}$ can be written down in a slightly more compact manner
\begin{eqnarray}
 \av{Z_{(ab)}} &=& \NPm{2Z_{(ij)}\,\xi\UD{i}{,A}\,\xi\UD{j}{,B}}_{ab} \label{eqcgzab}\\
 \av{Z_{[ab]}} &=& \Nn{g(N)_{\leftasym{a}}}_{\rightasym{b}}, \label{eqcgzab2}
\end{eqnarray}
 in which we have used the time derivative of (\ref{eqqAB}) to substitute $2Z_{(ij)}\,\xi\UD{i}{,A}\,\xi\UD{j}{,B}$ for $\dot q_{AB}$.
Note that the combination $\NPm{2\xi\UD{i}{,A}\,\xi\UD{j}{,B}\,\cdot\,}$ plays in our formalism the role of coarse--graining operator for $Z_{(ij)}$, a
symmetric tensors of rank 2 on $\Sigma_t$.

\begin{remark}


Although the coarse--grained $\av{Z_{ab}}$ is a generalization of the Newtonian volume average $\av{Q_{ab}}$, it does not have a straightforward interpretation as a genuine average of the local $Z_{\mu\nu}$ over $C_t$. Equations (\ref{eqQsym}) and (\ref{eqQasym}) can be written as volume integrals, just like (\ref{eqQbdr}), if we extend the velocity field $v^a$ to the interior of $D_t$ in an arbitrary way, but since we did not define any mapping between the interiors of $C_t$ and
$D_t$, such an extension would not be related to $Z_{\mu\nu}$ inside $C_t$ in any obvious way. It seems therefore more appropriate to refer to $\av{Z_{ab}}$ as a coarse--grained rather than average quantity.  
\end{remark}

\subsection{Preferred time derivative operator}
If we want to derive the equation for the time derivative of $\av{Z_{ab}}$, we need yet another ingredient. The equation involves the time derivative of a tensor
and there are obviously infinitely many different ways to differentiate it. In the infinitesimal counterpart of the discussed equation (\ref{eqduZ}) the role of the preferred time derivative is played by the covariant derivative along $u^\mu$, coinciding with the Fermi derivative along the particle trajectories. 
In this subsection we will introduce its generalization to the coarse--grained tensors in the case of irrotational fluid ($\omega_{ij} = 0$), with an adapted 3+1 splitting for which $u^\mu$ is orthogonal to $\Sigma_t$, i.e. $N^i = 0, N=1$.

Let $T\UD{ab\dots}{cd\dots}(t)$  be a time dependent but spatially constant field on the Euclidean space. In Cartesian coordinates in which condition (\ref{eqgauge}) holds
we set the preferred derivative to be equal to the ordinary time derivative
\begin{eqnarray*}
 \Dd T\UD{ab\dots}{cd\dots} = \frac{\partial}{\partial t}\,T\UD{ab\dots}{cd\dots}.
\end{eqnarray*}
In a more general situation, in Cartesian coordinates $x^a$ in $\Ee$ in which the image of embedding $f_t$ rotates one can prove that
\begin{eqnarray*}
\Dd T\UD{ab\dots}{cd\dots} &=& \frac{\partial}{\partial t}\,T\UD{ab\dots}{cd\dots} + W\UD{a}{z}\,T\UD{zb\dots}{cd\dots} + W\UD{b}{z}\,T\UD{az\dots}{cd\dots} +\cdots + \nonumber \\  &&- W\UD{z}{c}\,T\UD{ab\dots}{zd\dots} - W\UD{z}{d}\,T\UD{ab\dots}{cz\dots} + \cdots, 
\end{eqnarray*}
where
\begin{eqnarray}
 W_{ab} = -\frac{1}{V_0}\,\int_{\partial D_t} v_{\leftasym{a}}n_{\rightasym{b}}\,\dd\sigma.\label{eqWab}
\end{eqnarray}
measures  the net rotation of $\partial D_t$.

\section{Single worldline limit}

We will now justify the definitions of $\av{Z_{ab}}$ and, if the dust is irrotational, the preferred time derivative given above by showing that in the limit of the tube shrinking to a single worldline we recover the
standard 4--velocity gradient and Fermi derivative.

\begin{figure}
 \includegraphics[scale=0.85]{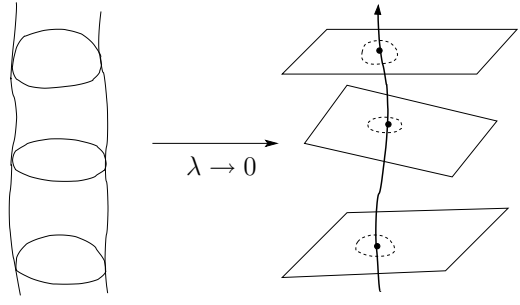}
 \caption{Single worldline limit of a tube}
 \label{figshrinking}
\end{figure}

Consider first a one parameter family of tubes given by 
\begin{eqnarray}
y^i = \lambda\,\xi^i(\theta^A). \label{eqlambda}
\end{eqnarray} 
where $\lambda$ is a positive parameter. In the limit of $\lambda \to 0$ the tube shrinks to a single worldline $y^i=0$ (see fig. \ref{figshrinking}). 
At a given instance of time $t= t_0$ the induced metric $q_{AB}$, expanded up to the subleading order terms in $\lambda$, has the form of
\begin{eqnarray}
 q_{AB}(t,\theta^C) = \lambda^2\,\xi\UD{i}{,A}\,\xi\UD{j}{,B}\,h_{ij}(0) + \lambda^3\,\xi\UD{i}{,A}\,\xi\UD{j}{,B}\,h_{ij,k}(0)
\,\xi^k(\theta^A) + O(\lambda^4) \label{eqqlambda}
\end{eqnarray}
where $h_{ij}(0)$ denotes the metric at point $p$ with coordinates $y^i = 0$.

Let $T_0$ denote the tangent space $T_p \Sigma_{t_0}$  to the point $p$ and let $w^i$ denote the canonical coordinates on it, related to $y^i$ by condition
$$I = w^i\,\pard{}{y^i}$$ for any vector $I \in T_0$. $T_0$ is a vector space, naturally endowed with a flat, Euclidean metric inherited from $\Sigma_t$. In the linear but most likely non--orthogonal coordinates $w^j$ 
the metric has the form of $h_{ij}(0)$. 

Since $T_0$ is a three--dimensional Euclidean space, we may use it as the target space for embeddings $f_{t_0}$ for all $\lambda > 0$ (see once again fig. \ref{figshrinking}).
In the leading order of $\lambda$ the embeddings must have the form of
\begin{eqnarray*}
 w^a &=& \chi^a(t,\theta^A,\lambda), 
\end{eqnarray*}
where $\chi^a$ can be expanded in terms of $\lambda$ as
\begin{eqnarray}
 \chi^a(t,\theta^A,\lambda) = \lambda\,\xi^a(\theta^A) + \frac{\lambda^2}{2}\,\Gamma\UD{a}{bc}(0)\,\xi^b(\theta^A)\,\xi^c(\theta^A) + O(\lambda^3), \label{eqembchi}
\end{eqnarray}
as usual up to rotations and translations. In order to prove it it is sufficient to calculate the metric induced on the image of embedding (\ref{eqembchi}), apply the formula for the Christoffel symbol and see that result agrees with
(\ref{eqqlambda}) up to $O(\lambda^4)$; the second part of theorem \ref{thiso} assures that (\ref{eqembchi})
 is unique up to rigid motions.

Note that since $T_0$ plays the role of the embedding space, we will refer to it by indices
$a,b,\dots$ rather that $i,j,\dots$, which we will reserve tangent spaces at other points.

\subsection{Expansion and shear}

We will now calculate $\av{Z_{(ab)}}$, in the leading order in $\lambda$, for the family of tubes defined above. 
Functions $\xi^i(\theta^A)$ do not depend on the coordinate time, but the metric $h_{ab}(0)$ at $p$ does. The time derivative of the induced metric 
is equal to
\begin{eqnarray}
 \dot q_{AB}(\theta^A) = \lambda^2\,\dot h_{ab}(t,0)\,\xi\UD{a}{,A}\,\xi\UD{b}{,B} + O(\lambda^3) \label{eqdotq}
\end{eqnarray}
in the leading order. The time derivative of the metric on the other hand is related to $Z\UD{i}{j}$ via
\begin{eqnarray}
 \dot h_{ij} = -2N\,K_{ij} + 2D_{\leftsym{i}}N_{\rightsym{j}} = 2Z_{(ij)} \label{eqdoth}.
\end{eqnarray}
in any comoving coordinate system.

Consider now the vector  field $V$ in $T_0$ given by
\begin{eqnarray*}
 V^a(w^b) = \frac{1}{2}\,\dot h_{bc}(0)\,h^{ac}(0)\,w^b + O(\lambda^2)
\end{eqnarray*}
($O(\lambda^2)$ denotes here a term which is of the order of $\lambda^2$ \emph{for fixed Euclidean coordinates $w^b$}, unlike in all other equations, 
where the limit $\lambda\to 0$ is taken \emph{for fixed boundary coordinates $\theta^A$}).
The restriction of $V^a$ to $\partial D_t$,
\begin{eqnarray}
 v^a(\theta^A) = \frac{\lambda}{2}\,\dot h_{bc}(0)\,h^{ac}(0)\,\xi^b(\theta^A) + O(\lambda^2), \label{eqvlinear}
\end{eqnarray}
induces the right change of metric in the leading order when the embedded surface is dragged along it:
\begin{eqnarray*}
 {\cal P}[v^a]_{AB} = \dot q_{AB} + O(\lambda^3).
\end{eqnarray*}
This can be proven by substituting (\ref{eqvlinear}) to (\ref{eqP1}) and comparison with (\ref{eqdotq}). From the second part of theorem \ref{thiso2}) we know that this is the only solution to (\ref{eqP2}) up to Euclidean motions. We can plug this guessed vector field into (\ref{eqQsym}) to obtain 
\begin{eqnarray*}
 \av{Z_{(ab)}} &=& \frac{1}{V_0}\,\int_{\partial D_t} v_{\leftsym{a}}\,n_{\rightsym{b}}\,\dd\sigma = \frac{1}{V_0}\,\int_{D_t}\,V_{(a,b)}\,\dd^3 x = \\
&=& \frac{1}{2}\,\dot h_{ab} + O(\lambda^2)
\end{eqnarray*}
or
\begin{eqnarray*}
 \av{Z_{(ab)}} = Z_{(ab)}(0) + O(\lambda^2).
\end{eqnarray*}

 \subsection{ Vorticity}

Calculating of the expansion and shear required only the retaining of the leading terms in $\lambda$ expansion. Vorticity requires including the next one as well,
because, as we shall see, it is the subleading term in $g(N)^a$ which contributes to $\av{Z_{[ab]}}$. 

For convenience we introduce auxiliary coordinates 
\begin{eqnarray*}
 z^i = y^i + \frac{1}{2}\,\Gamma\UD{i}{jk}(0)\,y^j\,y^k
\end{eqnarray*}
in the neighborhood of $p$ ($z^i = 0$). They agree with $y^i$ in the first order at the origin
\begin{eqnarray*}
 \pard{z^i}{y^j} = \delta\UD{i}{j}\Big|_{y^i = 0},
\end{eqnarray*}
but they are locally flat there in the sense that $h_{ij,k}(0) = 0$.
The coordinates $w^a$ on $T_0$ are also compatible with $z^j$, \emph{i.e.}
$I = w^a\,\pard{}{z^a}$ for any vector $I \in T_0$.

The tube is described in the auxiliary coordinates by equation
\begin{eqnarray*}
 z^i &=& \Xi^i(\theta^A) \\
 \Xi^i(\theta^A) &=& \lambda\,\xi^i(\theta^A) + \frac{\lambda^2}{2}\,\Gamma\UD{i}{kl}(0)\,\xi^k(\theta^A)\,\xi^l(\theta^A).
\end{eqnarray*}
This functional dependence is the same as in the equation  (\ref{eqembchi}) for isometric embedding in $T_0$:
\begin{eqnarray*}
 \chi^a(\theta^A) = \Xi^a(\theta^A) + O(\lambda^3).
\end{eqnarray*}

Consider the outward--pointing   normal to  $\partial C_t$,  given by
\begin{eqnarray*}
 \tilde n_k(\theta^A) &=& \frac{\pm\tilde b_k}{\sqrt{\tilde b_l\,\tilde b_m\,h^{lm}}} \\
 \tilde b_k(\theta^A) &=& \epsilon_{klm}\,\xi\UD{l}{,1}\,\xi\UD{m}{,2}
\end{eqnarray*}
($\epsilon_{ijk}$ stands for the totally antisymmetric symbol and the sign in the first equation depends on the orientation of $\theta^A$).
$\tilde n_k$, expressed in $z^i$, 
has the same functional dependence as the outward normal $n_a$ in $\Ee$, expressed in $w^b$, 
\begin{eqnarray*}
 n_a(\theta^A) = \tilde n_a(\theta^A) + O(\lambda^2) 
\end{eqnarray*}
up to the subleading order. This is a straightforward consequence of the fact that in the locally flat coordinates both the metric $h_{ij}(z^k)$ and the functions
$\Xi^i(\theta^A)$ agree with $h_{ab}(0)$ and $\chi^a(\theta^A)$ up to the subleading order.
Thus the pushforward of the shift vector agrees in these coordinates up to the subleading order as well:
\begin{eqnarray*}
 g(N)^a = N^a(0) + N\UD{a}{,b}(0)\,\lambda\,\xi^b(\theta^A) + O(\lambda^2).
\end{eqnarray*}
Operator ${\cal N}[\cdot]$ kills the constant part of $g(N)^a$ and decreases the order of other two terms by one, because the $\dd\sigma$ is of the order of $\lambda^2$, while the volume
$V_0$ of the order of $\lambda^3$:
\begin{eqnarray*}
 \av{Z_{[ab]}}={\cal N}\left[g(N)_{\leftasym{a}}\right]_{\rightasym{b}} = N_{[a,b]}(0) + O(\lambda)
\end{eqnarray*}
Since the covariant derivative $D_a$ is torsion free, this means that
\begin{eqnarray*}
 \av{Z_{[ab]}} = -D_{\leftasym{a}}N_{\rightasym{b}} + O(\lambda) = Z_{[ab]}(0) + O(\lambda).
\end{eqnarray*}

\subsection{Preferred time derivative operator}

Once again we assume in this subsection that the fluid does not rotate and $N^i = 0$. Let $T\UD{\mu\nu\dots}{\rho\sigma\dots}(t)$ be a family of tensors at tangent spaces at points crossed by the geodesic $y^i = 0$. We assume that the tensor field
is orthogonal to $u$ in each index. With this assumption $T$ can be reconstructed unambiguously from its projection to subspaces
$T_p\Sigma_t \subset T_p M$. Such tensors are effectively three--dimensional objects.

Consider the  projection of $\nabla_u T$ to $T_p\Sigma_t$ along $u^\mu$
\begin{eqnarray*}
\tDd T\UD{ab\dots}{cd\dots}=\nabla_u T\UD{\mu\nu\dots}{\rho\sigma\dots}\,\Pi\UD{a}{\mu}\,\Pi\UD{b}{\nu}\,\cdots \Pi\UD{\rho}{c}\,\Pi\UD{\sigma}{d}\cdots, 
\end{eqnarray*}
where $\Pi\UD{\alpha}{\beta} = \delta\UD{\alpha}{\beta} + u^\alpha\,u_\beta$. 
In the ADM variables this derivative is equal to
\begin{eqnarray*}
\tDd T\UD{ab\dots}{cd\dots} &= & \pard{}{t} T\UD{ab\dots}{cd\dots} + T\UD{pb\dots}{cd\dots}\,Z\UD{a}{p} + 
 T\UD{ap\dots}{cd\dots}\,Z\UD{b}{p} + \cdots\\
&-&  T\UD{ab\dots}{pd\dots}\,Z\UD{p}{c} - T\UD{ab\dots}{cp\dots}\,Z\UD{p}{d} - \dots. 
\end{eqnarray*}
 We will now show that $\Dd$ coincides with the $\tDd$ in the leading, zeroth order in $\lambda$.

Since $\Dd$ has been defined in orthonormal coordinates, we first need to introduce them on $T_0$. Let 
\begin{eqnarray*}
 x^a = \Lambda\UD{a}{b}\,w^b
\end{eqnarray*}
be such coordinates at $t = t_0$, \emph{i.e.} let
\begin{eqnarray*}
 \Lambda\UD{c}{a}\,\Lambda\UD{d}{b}\,\delta_{cd} = h_{ab}(0).
\end{eqnarray*}
We can extend them to other times by solving the initial value problem
\begin{eqnarray*}
 L\UD{a}{b}(t_0) &=& \Lambda\UD{a}{b}
\end{eqnarray*}
 for the ordinary differential equation
\begin{eqnarray}
 \dot L\UD{a}{b} &=& \frac{1}{2}\dot h_{bc}\,\left(L^{-1}\right)\UD{c}{d}\,\delta^{ad} \label{eqdL}
\end{eqnarray}
and taking $x_t^a = L\UD{a}{b}(t)\,w^b$ at all times. 

The antisymmetrized operator $\cal N$ vanishes when applied to the velocity field (\ref{eqvlinear}), so in the introduced coordinates the matrix $W_{ab}$ defining $\Dd$ 
has the form of
\begin{eqnarray*}
 W_{ab} = 0 + O(\lambda).
\end{eqnarray*}
Any tensor field expressed in $x^a$ can be converted back to $w^a$ by the formula
\begin{eqnarray}
 T\UD{ab\dots}{cd\dots} = \!^{(x)}T\UD{pq\dots}{rs\dots}\,\left(L^{-1}\right)\UD{a}{p}\,\left(L^{-1}\right)\UD{b}{q}\cdots L\UD{r}{c}\,L\UD{s}{d}\cdots. \label{eqxtow}
\end{eqnarray}
If we apply (\ref{eqxtow}) to calculate $\widetilde \Dd T\UD{ab\dots}{cd\dots}$ and make use of (\ref{eqdL}), we obtain 
after simplification the equation for $\Dd$ in $w^i$ coordinates
\begin{eqnarray*}
 \Dd T\UD{ab\dots}{cd\dots} = \tDd T\UD{ab\dots}{cd\dots} + O(\lambda).
\end{eqnarray*}

\section{Evolution equation for $\av{Z_{ab}}$}

In this section we will derive the evolution equation for $\av{Z_{ab}}$ and compare it with the evolution equation for $Z_{ij}$. We will only be concerned with
irrotational dust, leaving the case of non--vanishing rotation to the next paper.

In a spacetime filled with irrotational dust we can introduce comoving and orthogonal coordinates, in which $N=1$ and $N^i = 0$. 
Let $C_t$ be a fixed, comoving domain, whose boundary is admissible at least for some time. 
We can introduce a one--parameter family of isometric embeddings satisfying (\ref{eqgauge}) and assign the coarse--grained shear and expansion to it, encoded in $\av{Z_{ab}} = \av{Z_{(ab)}}$. 
This allows for decomposition of the \emph{local} $Z_{ab}$, pushed forward to $\partial D_t \subset \Ee$ via $g$, and the velocity field $v^a$ into the the coarse--grained, large--scale part and local inhomogeneities
\begin{eqnarray}
 Z_{ab} &=& \av{Z_{ab}} + \delta Z_{ab} \label{eqdeco1}\\
 v^a &=& \av{Z\UD{a}{b}}\,x^b + \delta v^a \label{eqdeco2},
\end{eqnarray}
both defined on $\partial D_t$ only.

We now pass to evaluating $\Dd \av{Z_{ab}}$.
We first note two identities concerning the time derivatives of operators $\Nn{\cdot}$ and $\Pp$ associated with $\partial D_t$. The first one,
\begin{eqnarray}
 \pard{}{t}\,\Nn{X_a}_b &=& -\av{Z\UD{c}{c}}\,\Nn{X_a}_b + \Nn{\frac{D X_a}{\partial t}}_{b} + \nonumber \\
&+&\Nn{v\UD{c}{,c}\,X_a}_b -
  \Nn{v\UD{c}{,b}\,X_a}_c, \label{eqident1}
\end{eqnarray}
is valid for any vector field $X^a$ defined on $\partial D_t$ for some time interval. Differential symbol $\frac{D}{\partial t}$ stands here
for the substantial derivative $\pard{}{t} + v^a\,\pard{}{x^a}$ in $\Ee$ along the velocity $v^b$, it should not be confused with the covariant derivative on $\Sigma_t$.
The second identity concerns the time derivative of $\Ppm{\,\cdot\,}$. Let $Y^a = \Ppm{r_{AB}}^a$, then
\begin{eqnarray}
 \frac{D Y_a}{\partial t} &=& \Ppm{\dot r_{AB}}_a - \Ppm{2v\UD{c}{,A}\,v_{c,B}}_a + \nonumber \\
 &+& \Nn{v\UD{c}{,c}\,Y_{\leftasym{a}}}_{\rightasym{b}}\,x^b - \Nn{v\UD{c}{\leftasym{,a}}\,Y_{\rightasym{b}}}_c\,x^b +
W_a \label{eqident2}
\end{eqnarray}
($W_a$ denotes, once again, an irrelevant constant vector, whose value depends on the condition we imposed on $\Ppm{\,\cdot\,}$). Both can be proven easily by taking the definitions of $\Nnn$ and $\Ppmm$, differentiating them with respect to time and rearranging the terms.

By applying (\ref{eqident1}) to the velocity field $v^a$ itself, we get
\begin{eqnarray}
 \pard{}{t}\Nn{v_a}_b &=& -\Nn{v^c}_c\,\Nn{v_b}_c + \Nn{\frac{D v_a}{\partial t}}_b + \nonumber\\ 
 &+&\Nn{v\UD{c}{,c}\,v_{a}}_b - \Nn{v\UD{c}{,b} \,v_a}_b. \label{eqdtn}
\end{eqnarray}
The convective derivative of $v^a$ can be related to the second derivative of the metric $q_{AB}$ using (\ref{eqident2}) 
\begin{eqnarray}
 \Nn{\frac{D v_a}{\partial t}}_b &=& \NPm{\ddot q_{AB}}_{ab} - \NPm{2 v\UD{c}{,A}\,v\UD{d}{,B}\,\delta_{cd}}_{ab} - \nonumber \\
&+&\Nn{v\UD{c}{,c}\,v_{\leftasym{a}}}_{\rightasym{b}} + \Nn{v\UD{c}{\leftasym{,a}}\,v_{\rightasym{,b}}}_c.
\label{eqdtp}
\end{eqnarray}
$\ddot q_{AB}$ is in turn related to the time derivative of $Z_{ij}$
\begin{eqnarray}
 \ddot q_{AB} = 2\dot Z_{ij}\,y\UD{i}{,A}\,y\UD{j}{,B} \nonumber
\end{eqnarray}
in ADM variables $(t,y^i)$.
From (\ref{eqduZ}) and (\ref{eqdoth}) we evaluate this derivative as
\begin{eqnarray}
 \dot Z_{ij} &=& Z_{ik}\,Z\UD{k}{j} - R_{i0j0} \label{eqriojo}
\end{eqnarray}
and thus, putting (\ref{eqdtn}), (\ref{eqdtp}), (\ref{eqchi}) and (\ref{eqriojo}) together,
\begin{eqnarray*}
 \pard{}{t} \Nn{v_a}_b &=& -\Nn{v^c}_c\,\Nn{v_a}_b + \NPm{2Z_{ce}\,Z\UD{e}{d}\,\chi\UD{c}{,A}\,\chi\UD{d}{,B}}_{ab} + \nonumber \\
&-& \NPm{2 v\UD{c}{\leftsym{,A}}\,v\UD{d}{\rightsym{,B}}\,\delta_{cd}}_{ab} - \NPm{2R_{i0j0}\,\xi\UD{i}{,A}\,\xi\UD{j}{,B}}_{ab} + \nonumber \\
&+& \Nn{v\UD{c}{,c}\,v_{\leftsym{a}}}_{\rightsym{b}} - \Nn{v\UD{c}{\leftsym{,a}}\,v_{\rightsym{b}}}_{c}.
\end{eqnarray*}
The left hand side of this equation is just the preferred time derivative $\Dd \av{Z_{ab}}$, because $N_a = 0$ and $\Nn{v_{\leftasym{a}}}_{\rightasym{b}} = 0$.
It now suffices to plug in the decomposition (\ref{eqdeco1})--(\ref{eqdeco2}) into the right hand side and rearrange the terms to obtain finally the evolution equation 
\begin{eqnarray}
 \Dd \av{Z_{ab}} = -\av{Z_{ac}}\,\av{Z\UD{c}{b}} - \av{R_{a0b0}}
 + B_{ab} + \widetilde B_{ab},
\label{eqdQ}
\end{eqnarray}
where 
\begin{eqnarray*}
 \av{R_{a0b0}} =  \NPm{2R_{i0j0}\,\xi\UD{i}{,A}\,\xi\UD{j}{,B}}_{(ab)}
\end{eqnarray*}
is obviously the coarse--graining of the contraction of the Riemann tensor with $u^\mu$, considered as a symmetric tensor on $\Sigma_t$ (compare with (\ref{eqcgzab})).

 This is clearly the coarse--grained counterpart of equation (\ref{eqduZ}). It includes two additional backreaction terms;
the first one is the symmetrized version of (\ref{eqnewtonbackreaction}), and we will call it the Newtonian backreaction 
\begin{eqnarray*}
 B_{ab} &=& \Nn{\delta v\UD{c}{,c}\,\delta v_{\leftsym{a}}}_{\rightsym{b}} - \Nn{\delta v\UD{c}{\leftsym{,a}}\,\delta v_{\rightsym{b}}}_c.
\end{eqnarray*}
The second one is entirely relativistic and has a more  complicated structure
\begin{eqnarray*}
 \widetilde B_{ab} &=& \NPm{4\av{Z_{cd}}\,\left(\delta Z\UD{d}{e}\,\chi\UD{e}{\leftsym{,A}} - 
 \delta v\UD{d}{\leftsym{,A}}\right)\,\chi\UD{c}{\rightsym{,B}}}_{ab} + \nonumber \\
 &+&\NPm{2\left(\delta Z_{ce}\,\delta Z\UD{e}{d}\,\chi\UD{c}{,A}\,\chi\UD{d}{,B} - \delta v\UD{c}{,A}\,\delta v\UD{d}{,B}\,
\delta_{cd}\right)}_{ab}.
\end{eqnarray*}
 Just like in the Newtonian case, both backreaction terms are surface integrals divided by volume. In contrast
to (\ref{eqnewtonbackreaction}), however, $\widetilde B_{ab}$ involves linear terms in perturbations. 

\section{Application to exact solutions}

We will now discuss briefly the application of the coarse--graining procedure to well--known exact solution of the Einstein  equations with dust. In particular,
we will show that
in case of metrics where matter undergoes homogeneous expansion or shear the coarse--grained quantities behave in a predictable way, \emph{i.e.} they are
equal to their spatially constant, local counterparts irrespective of the domain under consideration.

Consider first a FLRW metric with dust, closed, open or flat, in standard 3+1 decomposition
\begin{eqnarray}
 g &=& -\dd t^2 + a(t)^2\,h(t)  \label{eqflrw}\\
 h &=& \frac{\dd r^2}{1 - k\,r^2} + r^2\,\left(\dd \theta^2 + \sin^2\theta\,\dd \varphi^2\right) \nonumber.
\end{eqnarray}
 We will show that for any domain $C_t \subset \Sigma_t$, admissible in the sense of theorem \ref{thiso}, the coarse--grained
velocity gradient consists only of the scalar part, equal to 3 times the Hubble parameter
\begin{eqnarray}
 \av{Z_{ab}} = \frac{\dot a}{a}\,\delta_{ab} = H(t)\,\delta_{ab}. \label{eqplapla}
\end{eqnarray}
 The vanishing of the vorticity follows obviously from the vanishing of shift in (\ref{eqflrw}). The metric induced on the boundary of any domain in $\Sigma_t$ undergoes a homogeneous expansion
with time 
 \begin{eqnarray*}
  q_{AB}(t) = q_{AB}(t_0)\,\frac{a(t)^2}{a(t_0)^2},
 \end{eqnarray*}
whose momentary expansion rate is related to the Hubble parameter by
\begin{eqnarray*}
  \dot q_{AB}(t) = \frac{2\dot a}{a}\,q_{AB}(t) = 2H(t)\,q_{AB}.
\end{eqnarray*}
It is straightforward to see that the same change of metric is induced
on the image of the surface $\partial D_t$ in $\Ee$ by a homogeneous expansion along the vector field
\begin{eqnarray*}
 V^a = H(t)\,x^a,
\end{eqnarray*}
restricted to $\partial D_t$. (\ref{eqplapla}) follows now from (\ref{eqQsym}) if we use the divergence theorem to  turn it into a volume integral. 
In a similar fashion one can prove that in Kasner solutions \cite{Kasner} 
\begin{eqnarray*}
 \av{Z_{ab}} = \sigma_{ab} + \frac{1}{3}\,\theta\,\delta_{ab},
\end{eqnarray*}
where $\theta$ and $\sigma_{ab}$ are equal to the local expansion and shear. 

It is well--known that  it is possible to replace one or more non--intersecting balls within a dust FLRW solution by balls excised from 
an appropriately chosen LTB, Schwarzschild or Szekeres solution (so called Swiss cheese models, see \cite{Einstein-Strauss1, Einstein-Strauss2}, \cite{Krasinski}). In this case $\av{Z_{ab}}$ also consists only
of expansion, related to the Hubble constant of the exterior FLRW solution as in (\ref{eqplapla}), \emph{as long as the boundary $\partial C_t$ does not pass through any
of the excised balls} (though they can be present in any number in the \emph{interior} of the domain). This follows from the result above and from the fact that the presented coarse--grained quantities are quasi--local and thus insensitive to the changes of the
metric inside or outside $\partial C_t$. Thus in swiss cheese models our procedure recovers for appropriately chosen but generic domains the cosmological Hubble parameter of the background homogeneous solutions. 

The only metric in which dust undergoes homogeneous, rigid rotation, without expansion or shear, is the well--known G\"odel solution \cite{Goedel} with cosmological constant.
As one might expect, in this spacetime only the antisymmetric part of $\av{Z_{ab}}$ does not vanish, though its exact value depends on the 3+1 splitting.

A more detailed discussion of the application of the coarse--graining procedure to these and other exact solutions we be given in subsequent papers.

\section{Concluding remarks and acknowledgements}

In the subsequent papers we will derive the expression for backreaction for rotating dust and discuss more thoroughly its application to exact cosmological solutions. 

 The author would like to thank Lars Andersson, David Wiltshire, Michael Reiris and Gerhard Huisken for useful discussions and comments. The work was supported by the Foundation for Polish Science through the ''Master'' programme.

\bibliography{cosmo-paper-7}

\end{document}